\documentclass[aps,prd,twocolumn,preprintnumbers,floatfix,nofootinbib,notitlepage,superscriptaddress,showkeys,showpacs]{revtex4-1}

\usepackage{graphicx}
\usepackage{amsmath,amsfonts,amssymb}
\usepackage[hidelinks]{hyperref}
\usepackage{enumitem}
\usepackage{color}

\newcommand{\be}{\begin{equation}}
\newcommand{\ee}{\end{equation}}
\newcommand{\td}{{\rm d}}
\newcommand{\Msun}{M_{\odot}}

\begin{document}

\title{Primordial black hole constraints for extended mass functions}

\author{Bernard Carr}
\email{b.j.carr@qmul.ac.uk}
\affiliation{Astronomy Unit, Queen Mary University of London, Mile End Road, London, E1 4NS, U.K.}
\author{Martti Raidal}
\email{martti.raidal@cern.ch}
\affiliation{NICPB, R\"avala 10, 10143 Tallinn, Estonia}
\author{Tommi Tenkanen}
\email{t.tenkanen@qmul.ac.uk}
\affiliation{Astronomy Unit, Queen Mary University of London, Mile End Road, London, E1 4NS, U.K.}
\author{Ville Vaskonen}
\email{ville.vaskonen@kbfi.ee}
\affiliation{NICPB, R\"avala 10, 10143 Tallinn, Estonia}
\author{Hardi Veerm\"ae}
\email{hardi.veermae@cern.ch}
\affiliation{NICPB, R\"avala 10, 10143 Tallinn, Estonia}

\begin{abstract}
We revisit the cosmological and astrophysical constraints on the fraction of the dark matter in primordial black holes (PBHs) with an extended mass function. We consider a variety of mass functions, all  of which are described by three parameters: a characteristic mass and width and a dark matter fraction. Various observations then impose constraints on the dark matter fraction as a function of the first two parameters. We show how these constraints relate to those for a monochromatic mass function, demonstrating that they usually become more stringent in the extended case than the monochromatic one. Considering only the well-established bounds, and neglecting the ones that depend on additional astrophysical assumptions, we find that there are three mass windows, around $4\times 10^{-17}\Msun,$ $2\times 10^{-14}\Msun$ and $25-100 \Msun$, where PBHs can constitute all dark matter. However, if one includes all the bounds, PBHs can only constitute of order $10\%$ of the dark matter.
\end{abstract}

\maketitle

%===============================================================================
% BODY
%===============================================================================

%-------------------------------------------------------------------------------
\section{Introduction}
\label{sec:Introduction}
%-------------------------------------------------------------------------------

Besides its gravitational interaction, little is known about the nature of dark matter (DM) except that it is dynamically ``cold''. Although the cold dark matter (CDM) is usually assumed to be some form of elementary particle~\cite{Jungman:1995df,Bertone:2004pz}, there is still no evidence for this and PBHs which are too large to have evaporated by now are a possible alternative~\cite{Axelrod:2016nkp}. Because they form when the baryons only comprise a small fraction of the total cosmological density~\cite{Hawking:1971aa,Carr:1974nx,1975ApJ...201....1C}, they are exempt from the Big Bang nucleosynthesis limits on the baryonic density~\cite{CHAPLINE:1975aa}. Being much more  massive than elementary particles, they could also have a greater variety of observational consequences. Indeed  the PBH scenario is already severely constrained by cosmological and astrophysical observations~\cite{Carr:2009jm,Carr:2016drx}. 

The recent detection of gravitational waves from merging black holes with mass ${\mathcal O}(10)\Msun$ by LIGO~\cite{Abbott:2016blz,Abbott:2016nmj} has revived interest in the possibility of PBH DM~\cite{Kashlinsky:2016sdv,Bird:2016dcv,Clesse:2016vqa,Sasaki:2016jop}. Although the PBH coalescence rate depends on very uncertain astrophysical parameters, explaining the observed event rate would require the PBHs to contain at least a substantial fraction of the DM.

This has led to a reassessment of the existing PBH bounds in two directions. First, it has been argued that the existing constraints on PBHs with monochromatic mass functions can be relaxed by invoking extended mass functions~\cite{Clesse:2015wea,Inomata:2017okj}, the latter arising naturally if the PBHs are created from  inflationary fluctuations ~\cite{Carr:1994ar,GarciaBellido:1996qt,Kawasaki:1997ju,Yokoyama:1998pt,Kohri:2007qn,Frampton:2010sw,Drees:2011hb,Clesse:2015wea,Garcia-Bellido:2017mdw,Orlofsky:2016vbd,Inomata:2017okj,Kannike:2017bxn} or some form of cosmological phase transition~\cite{Crawford:1982yz,Hawking:1982ga,Khlopov:2008qy,Belotsky:2014kca}. However, there is still no rigorous treatment of how to apply the PBH bounds for an extended mass function and different analyses have led to different conclusions. For example, Ref.~\cite{Carr:2016drx} concludes that intermediate mass PBHs could provide the DM, whereas Refs.~\cite{Green:2016xgy} and \cite{Kuhnel:2017pwq} reach the opposite conclusion. Second, there have been revisions to  the constraints themselves, with some previous bounds being weakened (e.g.  those associated with accretion~\cite{Ricotti:2007au,Chen:2016pud,Ali-Haimoud:2016mbv,Blum:2016cjs}) and some new bounds being added. Indeed, the constraints are being constantly revised and the recent review of Ref.~\cite{Carr:2016drx} already needs to be updated. 

Currently there is no comprehensive study which combines these two approaches. In this paper, we fill this gap by presenting a general method for analysing the latest PBH constraints over the broad mass range $10^{-18}-10^{4} \Msun$ and applying them to an extended PBH mass function. 

It should be stressed that PBHs could play an important cosmological role even if they have much less than the DM density. For example, they could be useful  in explaining the rapid structure formation at small cosmological scales, provide seeds for supermassive black holes or galaxies and explain other unsolved astrophysical and cosmological puzzles~\citep{Garcia-Bellido:2017fdg,carrsilk}. This underlines the importance of knowing how the PBH density is distributed between different masses.

%-------------------------------------------------------------------------------
\section{Constraints on monochromatic PBH mass function}
\label{sec:constraints}
%-------------------------------------------------------------------------------

\begin{figure*}
\includegraphics[width=.42\textwidth]{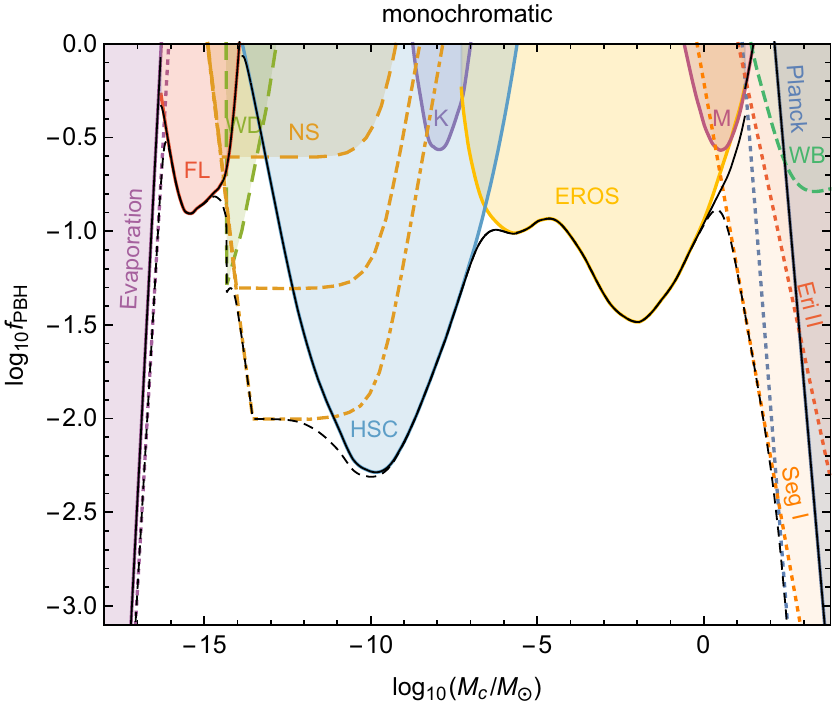} \hspace{6mm}
\includegraphics[width=.42\textwidth]{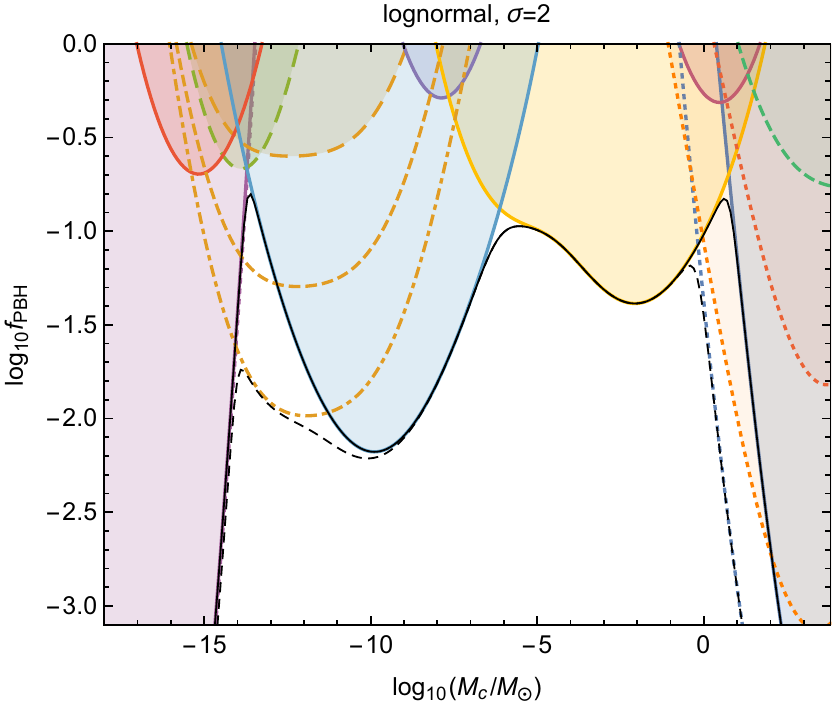} \\ \vspace{3mm}
\includegraphics[width=.42\textwidth]{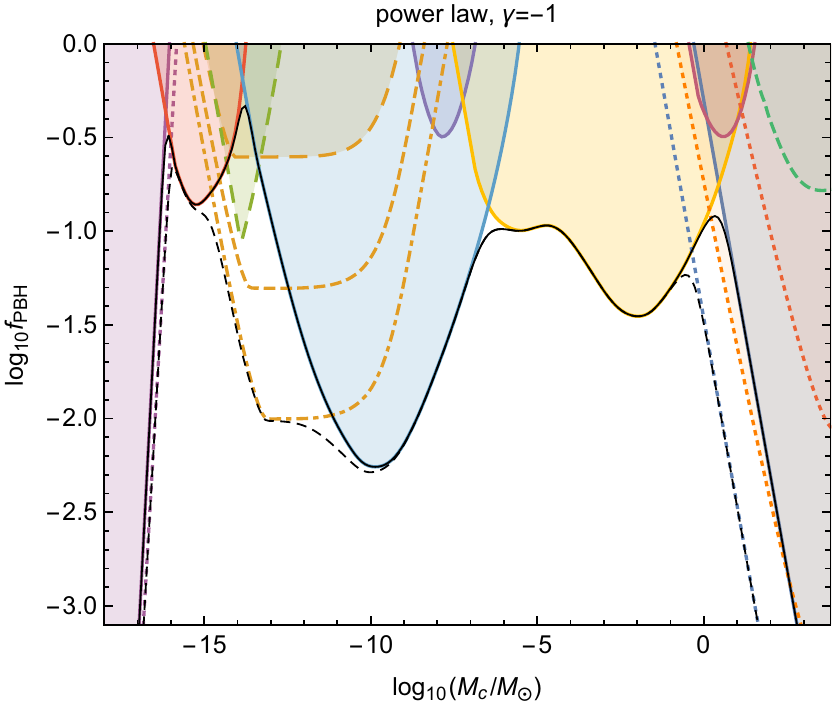} \hspace{6mm}
\includegraphics[width=.42\textwidth]{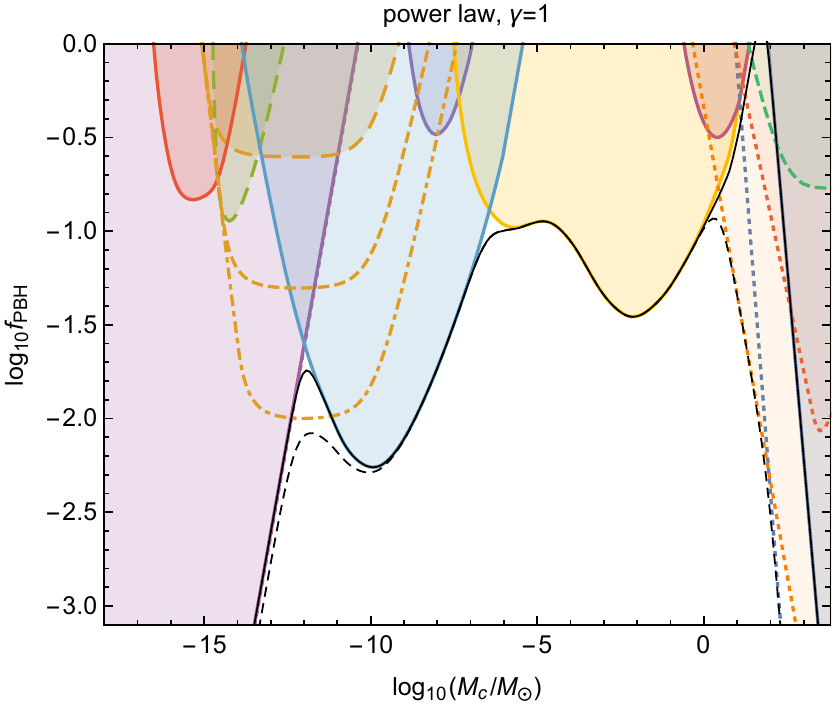}
\caption{\emph{Upper left panel:} Constraints from different observations on the fraction of PBH DM, $f_{\rm PBH}\equiv \Omega_{\rm PBH}/\Omega_{\rm DM}$,  as a function of the PBH mass $M_c$, assuming a monochromatic mass function. The purple region on the left is excluded by evaporations~\cite{Carr:2009jm}, the red region by femtolensing of gamma-ray bursts (FL)~\cite{Barnacka:2012bm}, the brown region by neutron star capture (NS) for different values of the dark matter density in the cores of globular clusters~\cite{Capela:2013yf}, the green region by white dwarf explosions (WD)~\cite{Graham:2015apa}, the blue, violet, yellow and purple regions by the microlensing results from Subaru (HSC)~\cite{Niikura:2017zjd}, Kepler (K) ~\cite{Griest:2013aaa}, EROS~\cite{Tisserand:2006zx} and MACHO (M)~\cite{Allsman:2000kg}, respectively. The dark blue, orange, red and green regions on the right are excluded by Planck data~\cite{Ali-Haimoud:2016mbv}, survival of stars in Segue I (Seg I)~\cite{Koushiappas:2017chw} and Eridanus II (Eri II)~\cite{Brandt:2016aco}, and the distribution of wide binaries (WB)~\cite{Monroy-Rodriguez:2014ula}, respectively.  The black dashed and solid lines show, respectively, the combined constraint with and without the constraints depicted by the colored dashed lines. \emph{Other panels:} Same as the upper left panel but for a lognormal PBH mass function with $\sigma=2$ (upper right) and for a power-law PBH mass function with $\gamma=-1$ (lower left) and $\gamma=1$ (lower right).}
\label{mono}
\end{figure*}

The main constraints on a PBH population derive from PBH evaporations, various gravitational lensing experiments, neutron star capture, numerous dynamical effects, and PBH accretion. The form of these constraints for a monochromatic PBH mass function is indicated in the upper left panel of Fig.~\ref{mono}, together with  relevant references. It must be stressed that these constraints depend on various cosmological and astrophysical assumptions, as well as unknown black hole physics. We therefore list these uncertainties explicitly.

The constraints on PBH evaporation via Hawking radiation~\cite{Hawking:1974rv} depend on the observed extragalactic photon flux intensity, $I\propto E^{-1- \epsilon},$ where $E$ is the photon energy  and $\epsilon$ parametrizes  the spectral tilt~\cite{Carr:2016drx}. There is some uncertainty in this parameter, so we present our results for the two extreme cases, $\epsilon=0.4$ (solid purple line)~\cite{Strong:2004ry} and $\epsilon=0.1$ (dotted purple line)~\cite{Sreekumar:1997un}. 
 
The Cosmic Microwave Background (CMB) anisotropy constraints on PBH accretion are subject to uncertainties in the accretion process and its effect on the thermal history of the universe at early times. To account for this, we show the bounds for both collisional ionisation (solid dark blue line) and photoionisation (dotted dark blue line)~\cite{Ali-Haimoud:2016mbv}. Recently, another sort of accretion limit has been obtained in the mass range from a few to $10^7 \Msun$ on the grounds that PBH accretion from the interstellar medium should result in a significant population of X-ray sources \cite{Inoue:2017csr}. Indeed, several earlier papers have considered such a limit \cite{Carr:1979,Gaggero:2016dpq}. However, all these limits are very dependent on the accretion scenario and are therefore not shown. 
 
Lensing is the only phenomenon which has been claimed to provide {\it positive} evidence for PBHs. For example, the results of the MACHO project -- searching for microlensing of stars in the Magellanic clouds -- originally suggested halo DM in the form of $0.5\Msun$ objects~\cite{Alcock:1996yv} and these could plausibly be PBHs formed at the quark-hadron phase transition at $10^{-5}$s. However, the DM fraction was later reduced to 20\%~\cite{Alcock:2000ph}. The interpretation of the MACHO results -- and also the EROS and OGLE results -- is very sensitive to the properties of the Milky Way halo. In particular, it has been argued that the recent low-mass Galactic halo models would relax the constraints and allow the halo to consist entirely of solar mass PBHs~\cite{2015A&A...575A.107H}. Where only a constraint is claimed, rather than   a positive detection, it is important to specify the associated confidence level (CL). For all lensing constraints shown in Fig.~\ref{mono}, we use the 95\% CL constraint given in Refs.~\cite{Niikura:2017zjd,Griest:2013aaa,Tisserand:2006zx,Allsman:2000kg}.

There is also positive evidence for microlensing from quasar observations, these indicating that 20\% of the total mass is in compact objects in the mass range $0.05-0.45\Msun$~\cite{Mediavilla:2017bok}. This is compatible with the expected characteristics of the stellar component and the observations may also exclude a significant population of PBHs outside this mass range. However, this constraint is not used in our analysis because it is difficult to express the result of Ref.~\cite{Mediavilla:2017bok} as a quantitative upper bound on the PBH mass fraction. This conclusion may also conflict with recent claim that long-term radio variability in the light-curves of active galactic nuclei (AGN) arises from gravitational millilensing of features in AGN jets~\cite{Vedantham:2017kyb}. This claim could imply that the DM is either individual black holes of mass $10^3 - 10^6 \Msun$ or clusters of this mass comprising smaller black holes. 

In the latter context, additional relaxing of constraints would apply if the PBHs were spatially clustered into sub-haloes. As stressed in Ref.~\cite{Clesse:2016vqa}, this is expected if PBHs are part of a larger-scale overdense region. However, this effect depends on details of small-scale structure formation which are not fully understood, so we simply adopt the results presented in the current literature.

Observations of neutron stars limits the PBH abundance and indeed it has been claimed that this excludes PBH DM over a wide range of masses. However, these limits are dependent on the DM density in the cores of globular clusters, which is very uncertain. Following Ref.~\cite{Capela:2013yf}, the neutron star capture constraint is presented for three values of  this density (dashed and dot-dashed yellow lines).
 
It must be stressed that the constraints in Fig.~\ref{mono} have varying degrees of certainty and they all come with caveats. For some, the observations  are well understood (e.g. the CMB and gravitational lensing data) but there are uncertainties in the black hole physics. For others, the observations themselves are not fully understood or depend upon additional astrophysical assumptions. To address the associated uncertainties in a systematic way, we split the constraints into two classes. The first class, presented in Fig.~\ref{mono} by solid lines, are relatively robust, while the second class, presented by dashed lines, are somewhat less firm and depend upon astrophysical parameters. In particular, this applies to most of the dynamical and accretion constraints (e.g. those associated with dwarf galaxies, wide binaries and neutron stars). However, we stress that this division is not completely clear-cut. In the following, we present our results for the two classes of constraints both separately and together.

%-------------------------------------------------------------------------------
\section{Constraints on extended PBH mass function}
\label{sec:emf}
%-------------------------------------------------------------------------------

If the PBHs span an extended range of masses, the mass function is usually written as $\td n/\td M$ where $\td n$ is the number density of PBHs in the mass range $(M, M+\td M)$. For our purposes it is more convenient to introduce the function
\be
	\psi(M) \propto M \frac {\td n}{\td M} \, ,
\ee
normalised so that the fraction of the DM  in PBHs is
\be\label{eq:normalization}
	f_{\rm PBH} \equiv \frac{\Omega_{\rm PBH}}{\Omega_{\rm DM}} = \int \td M \,  \psi(M)  \, ,
\ee
where $\Omega_{\rm PBH}$ and $\Omega_{\rm DM}\approx 0.25$ are the PBH and DM densities in units of the critical density. The lower cut-off in the mass integral  necessarily exceeds $M_*  \approx 4 \times 10^{14}$g, the mass of the PBHs evaporating at the present epoch \cite{Carr:2009jm}. Note that $\psi(M)$ is the distribution function of $\log{M}$ and has units [mass]$^{-1}$.

In this paper we consider three types of mass function.

\begin{enumerate}[leftmargin=*]
\item A lognormal mass function of the form:
\be\label{dist}
	\psi(M) = \frac{f_{\rm PBH}}{\sqrt{2\pi}\sigma M} \exp\left(-\frac{\log^2(M/M_c)}{2\sigma^2}\right) \,,	
\ee
where $M_c$ is the mass at which the function $M\psi(M)$ peaks and $\sigma$ is the width of the spectrum. This was first suggested in Ref.~\cite{Dolgov:1992pu} and is often a good approximation if the PBHs result from a smooth symmetric peak in the inflationary power spectrum. This was demonstrated numerically in Ref.~\cite{Green:2016xgy} and analytically in Ref.~\cite{Kannike:2017bxn} for the case in which the slow-roll approximation holds. It is therefore representative of a large class of extended mass functions. Note that the lognormal mass function used in Refs.~\cite{Green:2016xgy,Horowitz:2016lib,Kuhnel:2017pwq} omits the $M^{-1}$ term in Eq.~\eqref{dist}.  In this case, the position of the peak of $M\psi(M)$ is no longer $M_c$ but $e^{\sigma^2}M_c$. The form \eqref{dist} is more useful for our purposes because  $M \psi(M)$  relates to the DM fraction in PBHs of mass $M$.

\item A power-law mass function of the the form
\be
	\psi(M) \propto  M^{\gamma -1} \quad (M_{\rm min} < M < M_{\rm max}) \, .
\label{power}
\ee 
For $\gamma\neq 0$, either the lower or upper cut-off can be neglected if $M_{\rm min}\ll M_{\rm max}$, so this scenario is effectively described by two parameters. Only in the $\gamma = 0$ case are both cut-offs necessary. For example, a mass function of this form arises naturally if the PBHs form from scale-invariant density fluctuations or from the collapse of cosmic strings. In both cases,  $\gamma = - 2w/(1+w)$, where $w$ specifies the equation of state, $p = w \rho$, when the PBHs form \cite{1975ApJ...201....1C}. In a non-inflationary universe, $w \in (-1/3,1)$ and so the natural range of the mass function exponent is $\gamma \in (-1,1)$. Equation~\eqref{power}  is not applicable for $w\in(-1,-1/3)$, corresponding to $\gamma\in(1,\infty)$, because PBHs do not form {\it during} inflation but only {\it after} it as a result of inflation-generated density fluctuations. Special consideration is also required in the  $w=0$ (matter-dominated) case~\cite{Khlopov:1980mg,Polnarev:1986bi}, because then both cut-offs in \eqref{power} can be  relevant and this is discussed elsewhere~\cite{Carr:2017}. In the following analysis we will consider both positive and negative values for $\gamma$ but not zero.

\item A critical collapse mass function~\cite{Yokoyama:1998xd,Niemeyer:1999ak,Musco:2012au,Carr:2016hva}:
\be
	\psi(M) \propto M^{2.85} \exp(-(M/M_f)^{2.85})\, ,
\label{crit}
\ee
which may apply generically if the PBHs form from density fluctuations with a $\delta$-function power spectrum. In this case, the mass spectrum extends down to arbitrarily low masses but there is an exponential upper cut-off at a mass-scale $M_f$ which corresponds roughly to the horizon mass at the collapse epoch. If the density fluctuations are themselves extended, as expected in the inflationary scenario,  then Eq.~\eqref{crit} must be modified  \cite{Carr:2016drx}. Indeed, the lognormal distribution may then be appropriate. So although the mass function \eqref{crit} is described by a single parameter, two may be required in the more realistic critical collapse situation.
\end{enumerate}

To compare with the lognormal case, we describe the mass function in the last two cases by the mean and variance of the $\log M$ distribution:
\be
	\log M_{c} \equiv \langle \log M \rangle_{\psi} , \quad
	\sigma^{2}  \equiv \langle \log^{2} M \rangle_{\psi} - \langle \log M \rangle_{\psi}^{2} \, ,
\ee
where $\langle X \rangle_{\psi} \equiv f_{\rm PBH}^{-1}\int \td M\, \psi(M) X(M)$. For a power-law distribution these are
\be
	M_{c}	= M_{\rm cut} e^{-\frac{1}{\gamma}}, \qquad
	\sigma 	= \frac{1}{|\gamma|} \,,
\label{par}
\ee
where $M_{\rm cut}$ stands for $ {\rm max} (M_{\rm min}, M_*)$ if $\gamma < 0$ or  $M_{\rm max}$ if $\gamma > 0$. For the critical-collapse distribution \eqref{crit}, the exponential cut-off is very sharp, so the mass function is well approximated by a power law distribution with $\gamma = 3.85$ and $M_{\rm max} \approx M_f$. As it is relatively narrow, Eq.~\eqref{par} implying $\sigma = 0.26$, even the monochromatic mass function provides a good fit. Since critical collapse should be a fairly generic feature of PBH formation, $\sigma = 0.26$ will usually provide a lower limit to the width of the mass function. However, critical collapse may not be relevant in all cases, for example in the cosmic string or matter-dominated ($w=0$) scenarios.

It should be stressed that two parameters should always suffice to describe the PBH mass function {\it locally} (i.e. close to a peak) since this just corresponds to the first two terms in a Taylor expansion. However, in principle the mass function could be more complicated than this. For example, depending on the form of the inflaton potential, it could have several distinct peaks. Indeed, with a sufficiently contrived form, these peaks could be tuned to exactly match all the constraint windows.

The existing constraints on the allowed fraction of PBH DM are commonly presented assuming a monochromatic mass function (presented in the upper panel of Fig.~\ref{mono}). In the following we introduce a simple method for generalising these results to arbitrary mass functions. For this purpose, consider an astrophysical observable $A[\psi(M)]$ depending on the PBH abundance (e.g. the number of microlensing events of given duration in a given time interval). It can generally be expanded as
\be
\label{eq:A_observable}
\begin{aligned}
	&A[\psi(M)] = A_0 + \int \td M\, \psi(M) K_{1}(M) \\
	&+ \int \td M_{1} \td M_{2}\, \psi(M_{1})\psi(M_{2}) K_{2}(M_{1},M_{2}) + \ldots ,
\end{aligned}
\ee
where $A_0$ is the background contribution and the functions $K_j$ depend on the details of the underlying physics and the nature of the observation. If PBHs of different mass contribute independently to the observable, only the first two terms in Eq.~\eqref{eq:A_observable} need to be considered. Explicit expressions are given for lensing and survival of stars in Ref.~\cite{Green:2016xgy}, evaporation in Ref.~\cite{Carr:2016drx}, and neutron star capture and accretion in Ref.~\cite{Kuhnel:2017pwq}. In this case, if a measurement puts an upper bound on the observable,
\be \label{eq:Aexp}
	A[\psi(M)]  \leq A_{\rm exp},
\ee
then for a monochromatic mass function with $M=M_c$,
\be
	\psi_{\rm mon}(M) \equiv f_{\rm PBH}(M_{c}) \delta(M - M_{c}),
\ee
this translates to 
\be\label{eq:f_max}
	f_{\rm PBH}(M_{c})  \leq \frac{A_{\rm exp} - A_0}{K_{1}(M_{c})} \equiv f_{\rm max}(M_{c}) \, .
\ee
The function $f_{\rm max}(M)$ corresponds to the maximum observationally allowed fraction of DM in PBHs for a monochromatic mass distribution. Combining Eqs.~(\ref{eq:A_observable})--(\ref{eq:f_max}) then yields
\be\label{eq:general_constraint}
	\int \td M  \frac{\psi(M)}{f_{\rm max}(M)} \leq 1 \, .
\ee
Once $f_{\rm max}$ is known, it is possible to apply Eq.~\eqref{eq:general_constraint} for an arbitrary mass function $\psi (M)$ to obtain the constraints equivalent to those for a monochromatic mass function. 

In detail the procedure is as follows. We first integrate Eq.~\eqref{eq:general_constraint} over the mass range ($M_1,M_2$) for which the constraint applies, assuming a particular function $\psi (M;f_{\rm PBH},M_c, \sigma)$. Once we have specified $M_1$ and $M_2$, this constrains $f_{\rm PBH}$ as a function of $M_c$ and $\sigma$. (In all cases except lensing, we take the integral limits to be the values of $M$ for which $f_{\rm max} =100$.) The last three panels in Fig.~\ref{mono} are then derived by assuming $\sigma=2$ for the lognormal mass function (upper right panel) and $\gamma=\pm1$ for the power law mass function (lower panels).

The procedure must be implemented separately for each observable. As shown in the \hyperlink{sec:combining}{Appendix}, different constraints can be combined by using the relation
\be\label{eq:combined_constraint}
	\sum_{j=1}^N \left(\int \td M \frac{\psi(M)}{f_{{\rm max},j}(M)} \right)^2 \leq 1 \,,
\ee
where $f_{{\rm max},j}(M)$ correspond to the different bounds for a monochromatic mass function, as defined by Eq.~\eqref{eq:f_max}. Most of the constraints shown in Fig.~\ref{mono} rely on a single observable. For lensing this is the number of lensing events~\cite{Barnacka:2012bm,Niikura:2017zjd,Tisserand:2006zx,Allsman:2000kg}, for neutron star capture it is the age of neutron stars~\cite{Capela:2013yf}, and for white dwarfs and wide binaries it is their abundance~\cite{Graham:2015apa,Monroy-Rodriguez:2014ula}.

However, some monochromatic constraints reported in the literature contain contributions from multiple observables. For example, consider the Planck constraint of Ref.~\cite{Ali-Haimoud:2016mbv}. Earlier analyses calculated the optical depth from the CMB data and used that to constrain the PBH abundance~\cite{Ricotti:2007au,Chen:2016pud}. In this case, there is only one observable, the optical depth, and Eq.~\eqref{eq:general_constraint} is applicable. However, the constraint from CMB anisotropies calculated in Ref.~\cite{Ali-Haimoud:2016mbv} combines experimental data at various multipole moments by performing a $\chi^2$ analysis for $f_{\rm PBH}$ using the CMB data. It is shown in the \hyperlink{sec:combining}{Appendix} that using the combined bound $f_{{\rm max},{\rm CMB}}(M)$ in Eq.~\eqref{eq:general_constraint}, instead of considering the contributions from different multipoles separately, will result in a more stringent constraint. However, since the mass dependence of the PBH contribution is expected to be roughly proportional for different multipoles~\cite{Ali-Haimoud:2016mbv}, the error should be small compared to the theoretical uncertainties, shown in Fig.~\ref{mono}, associated with PBH accretion physics. To accurately estimate the size of this error, one should repeat the analysis of Ref.~\cite{Ali-Haimoud:2016mbv} for each mass function separately, which is beyond the scope of this work.

The important qualitative point is that the form of Fig.~\ref{mono} in the non-monochromatic case is itself dependent on the PBH mass function. One cannot just compare a predicted extended mass function with the monochromatic form of the constraints, as some authors have done. In displaying the constraints, one also needs to select values of the parameters which describe the mass function. In both the lognormal and power-law cases, we have taken these to be $\sigma$ and $M_c$. For the critical collapse model, there is only one parameter ($M_f$) but this model is practically indistinguishable from the monochromatic one because only a small fraction of the PBH density is associated with the  low-mass tail. So this case is not shown explicitly. 

We now discuss some caveats that have to be kept in mind when applying Eq.~\eqref{eq:general_constraint}. The mass function evolves in time if the PBH merge or if new black holes are created. This can have an important impact on the constraints. For example, if mergers between recombination and the present are significant, the accretion constraints will be relaxed, since the mass function at recombination would have peaked at a lower mass than today. A period of merging after recombination is not implausible, as this would be induced by the small-scale density fluctuations which are likely to accompany PBH production~\cite{Hawking:1971aa,Carr:1974nx}. 

We next discuss the effect of the higher order terms in \eqref{eq:A_observable} since these can induce errors in Eq.~\eqref{eq:general_constraint}. These terms become relevant if the contribution of a black hole population with a given mass is influenced by the presence of another black hole population with a different mass. For example, the non-detection of a stochastic gravitational wave background from PBH binaries may constrain the PBH abundance in the near future~\cite{Wang:2016ana}. This constraint depends on the $K_2$ term in Eq.~\eqref{eq:A_observable} and it may also depend on $K_3$ if the formation of PBH binaries depends on $3$-body effects~\cite{Ioka:1998nz}.

In some cases, it is possible to remove the higher order terms by introducing an `effective' mass function. For example, compact gravitationally bound systems (such as binaries) may behave as a single objects in the context of lensing.  Consider an idealised scenario in which the observable depends purely on the total mass of the object, so that $K_{n}(M_{1},M_{2},\ldots, M_{n}) = K_{n}(M_{1}+M_{2}+\ldots + M_{n})$, but is otherwise independent of the composition, so that $K_{n} \propto K_{1}$. If we additionally assume that the mass function within these compact bound systems follows the overall mass mass function, we obtain
\be 
	A[\psi(M)] \approx A_0 + \int \td M\,  K_{1}(M) \psi_{\rm eff}(M) \, .
\ee
Here the effective mass function is given by
\be
	\psi_{\rm eff}(M) = \sum_{n}\alpha_{n} \psi_{n} (M) \, ,
\ee
where $\alpha_{n}$ relates to the fraction of $n$-body bound objects, 
\be
	\psi_{n} (M) \equiv \int \prod^{n}_{i=1} \td M_{i} \psi(M_{i})  \delta(M - \Sigma M_{i}) \, ,
\ee
and the effective mass function $\psi_{\rm eff}$ has to satisfy the normalisation condition $f_{\rm PBH}\leq 1$. The constraints for the general and monochromatic mass functions are still related by \eqref{eq:general_constraint} but likely overestimate the allowed PBH mass since $\psi_{\rm eff}(M)$ is always shifted towards higher masses. In principle, all the constraints discussed below and shown in our figures relate to the {\it effective} mass functions, which can be different for different constraints. 

\begin{figure*}
\includegraphics[width=\textwidth]{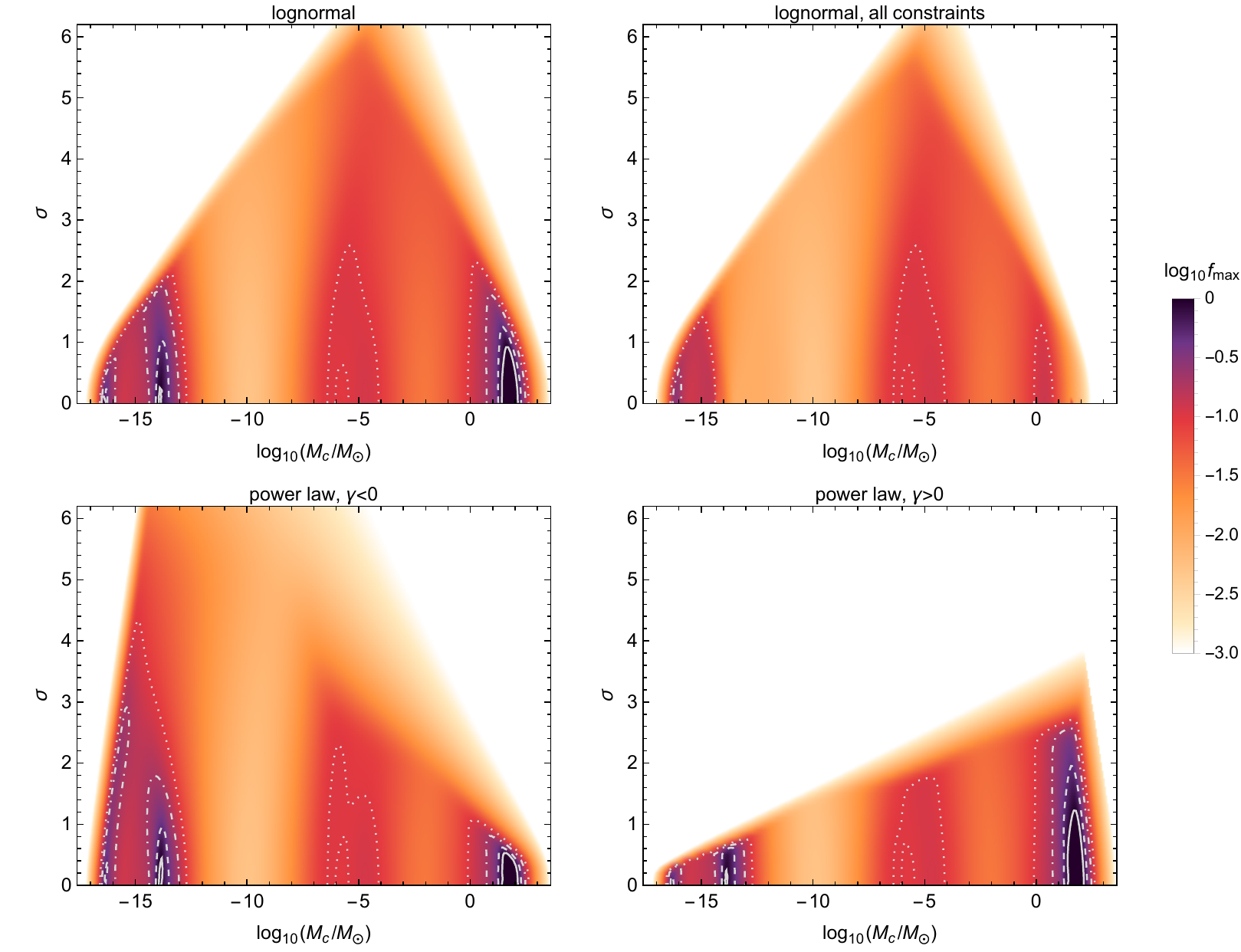}
\caption{\emph{Upper panels:} Combined observational constraints on $M_c$ and $\sigma$ for a lognormal PBH mass function. The color coding shows the maximum allowed fraction of PBH DM. In the white region $\log_{10}f_{\rm max}<-3$, while the solid, dashed, dot-dashed and dotted contours correspond to $f_{\rm max}=1$, $f_{\rm max}=0.5$, $f_{\rm max}=0.2$ and $f_{\rm max}=0.1$, respectively. In the left panel only the constraints depicted by the solid lines in Fig.~\ref{mono} are included, whereas the right panel includes all the constraints. \emph{Lower panels:} Same as the upper left panel but for a power-law mass function with $\gamma<0$ (left) and $\gamma>0$ (right).}
\label{densityplot}
\end{figure*}

It is also possible that the mass function is position-dependent. This is expected in dwarf galaxies because mass segregation causes lighter PBHs to migrate outwards, with the heavier ones occupying the central region. This will introduce corrections for constraints arising from the evolution of stars in the Galaxy \cite{Koushiappas:2017chw, Brandt:2016aco}. Again, it might be possible to invoke an effective mass function $\psi_{\rm eff}$ that only accounts for the heavier PBHs. However, an estimate of this effect requires detailed numerical simulations which are beyond the scope of this work. 

\begin{figure*}
\includegraphics[width=.42\textwidth]{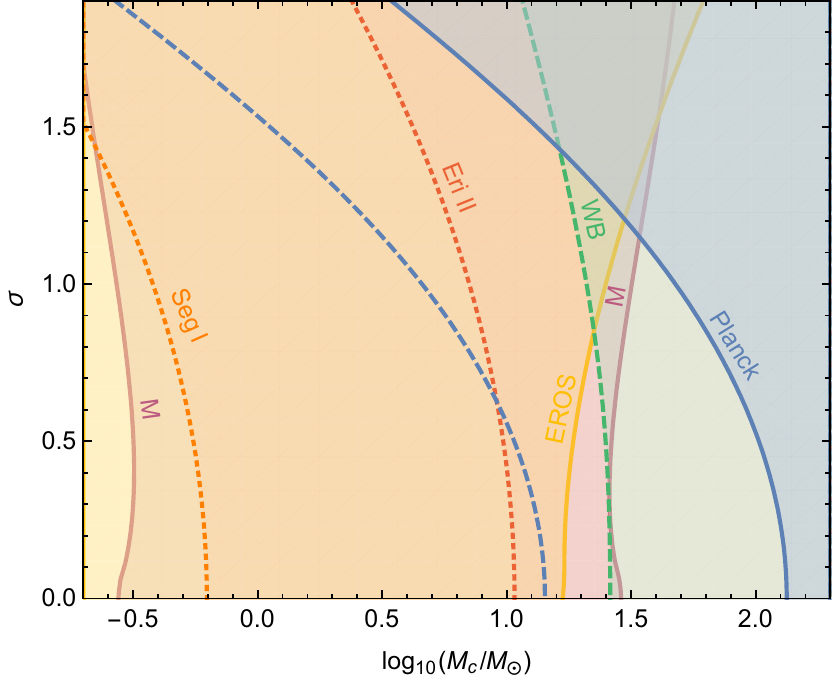} \hspace{6mm}
\includegraphics[width=.42\textwidth]{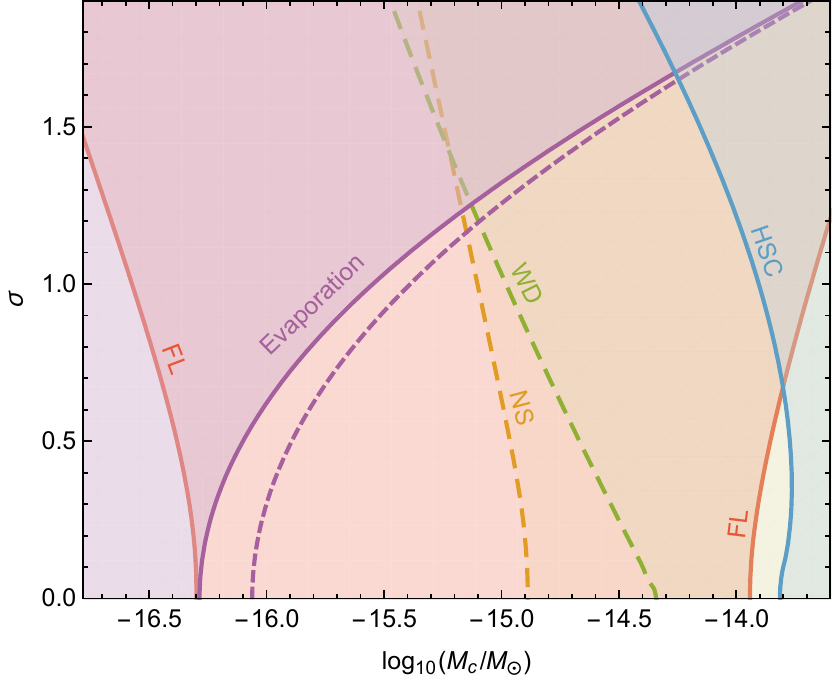}
\caption{Observational constraints on $M_c$ and $\sigma$ for a lognormal PBH mass function, assuming 100\% PBH DM. The left panel presents a zoom into the high mass region relevant for the LIGO events, while the right panel presents a zoom into the low mass region.
The color coding is the same as in Fig.~\ref{mono}.}
\label{fPBH1}
\end{figure*}

%-------------------------------------------------------------------------------
\section{Results and discussion}
\label{sec:results}
%-------------------------------------------------------------------------------

Our main results are presented in Fig.~\ref{densityplot}, where we show constraints on the maximum allowed fraction of PBH DM, $f_{\rm max}$, in the ($M_c, \sigma$) plane  for lognormal and power-law PBH mass functions. In the upper right panel all the constraints shown in Fig.~\ref{mono} are considered,  using the most restrictive forms for the evaporation, accretion and neutron star constraints, as depicted by the dotted lines. In the other panels only the constraints corresponding to the solid lines are taken into account. We have combined the constraints using Eq.~\eqref{eq:combined_constraint}. The black lines in Fig.~\ref{mono} correspond to constant $\sigma$ slices in Fig.~\ref{densityplot}. The regions where 10\%, 20\%, 50\% and 100\% of DM can consist of PBHs are indicated in Fig.~\ref{densityplot} by the dotted, dot-dashed, dashed and solid lines, respectively, while less than 0.1\% of the DM can be in PBHs in the white region.

The shape of the constraints in Fig.~\ref{densityplot} makes it clear that the allowed mass range for fixed $f_{\rm PBH}$ decreases with increasing the width $\sigma$, thus ruling out the possibility of evading the constraints by simply extending the mass function. Moreover, Fig.~\ref{densityplot} gives an upper bound $\sigma \lesssim 1$ if all dark matter is in the form of PBHs. This implies $|\gamma| \gtrsim 1$, which effectively rules out PBH DM from the collapse of cosmic strings or scale-invariant density fluctuations. 

Our results agree with the conclusions of Refs.~\cite{Green:2016xgy,Horowitz:2016lib,Kuhnel:2017pwq}. However, Refs.~\cite{Green:2016xgy,Horowitz:2016lib} focus on PBHs in the solar to intermediate mass range, considering microlensing and dynamical constraints from Eridanus II. Ref.~\cite{Kuhnel:2017pwq} performs a more comprehensive analysis, covering the mass range $10^{-18} - 10^4 \Msun$, but their study does not include the recent constraint from Subaru Hyper Suprime-Cam~\cite{Niikura:2017zjd} and they calculate the Planck constraint as in Ref.~\cite{Chen:2016pud},  resulting in a more stringent constraint than the one from Ref.~\cite{Ali-Haimoud:2016mbv} used in this work. Also they use the potential SKA pulsar timing constraints~\cite{Schutz:2016khr}, even though these are not yet realised. Some of the difference between our figures and those in Refs.~\cite{Green:2016xgy,Horowitz:2016lib,Kuhnel:2017pwq} results from the difference in the definition of $M_c$.

The same conclusion can be drawn if one compares the constraints presented in the upper left and right panels of Fig.~\ref{mono}. {In the latter case, we show the corresponding  $(f_{\rm PBH},M_c)$ constraints for extended mass functions with fixed width. The effect of the extension is to  `smooth' the constraints.  Although the most restrictive constraints for the PBH fraction are weakened, it can be seen that  the regions allowing a relatively large PBH fraction are reduced. So the constraints become wider, as indicated in Fig.~\ref{mono}. We conclude that previous claims in the literature that wide mass functions allow one to avoid PBH bounds are premature and not supported by our more rigorous computations.

The shape of the colored region of Fig.~\ref{densityplot} can be understood as follows: The lognormal mass function is symmetric in the $\log M$ scale, while the power law with $\gamma<0$ has a high-mass tail and $\gamma>0$ is skewed towards low masses. Since the evaporation constraint~\cite{Carr:2016drx} is much stronger than the accretion one~\cite{Carr:2009jm}, the low-mass tail excludes wider mass functions, whereas $\gamma<0$ allows it.

There are three regions in the upper left panel of Fig.~\ref{densityplot} where all DM can consist of PBHs. Two of them are at very low mass, just above the evaporation limit, and the third is  in the mass window relevant for the LIGO black hole coalescence events. However, this neglects the dynamical constraints, shown by the dashed lines in Fig.~\ref{mono}. As explained above, this might be justified for reasons associated with the dynamics of the observed astrophysical systems.

To clarify what role different constraints play in the regions of interest, we present these regions in detail in Fig.~\ref{fPBH1} for $\Omega_{\rm PBH}=\Omega_{\rm DM}$. The masses $25 - 100 \Msun$ satisfy the microlensing and accretion constraints but conflict with dynamical constrains from ultra-faint dwarfs and wide binaries. At the lower mass end, there is a narrow window around $4\times10^{-17}\Msun$ if we assume a conservative bound from evaporations and another window around $2\times10^{-14} \Msun$ if the dynamical constrains associated with neutron stars and white dwarfs are neglected. Both masses are in the asteroid range. If all the constraints are taken into account, the maximally allowed fraction of PBH DM is 13\%  in the high mass window and 30\% and 15\% in the two low mass windows, respectively. Note that whether the DM can be in PBHs in the asteroid window is sensitive to the form of the PBH evaporation limit and this depends on the precise form of the extragalactic $\gamma$-ray background.

%-------------------------------------------------------------------------------
\section{Conclusions} 
\label{sec:Conclusions}
%-------------------------------------------------------------------------------

We have studied the constraints on PBH DM with an extended mass function, presenting a general method for extracting these constraints from those for monochromatic PBH mass functions and discussing possible caveats associated with their interpretation. Our computations cover the broad mass range $10^{-18} - 10^{4}\Msun$ and show that extended mass functions do not generally alleviate the already existing constraints on the PBH DM fraction, because the allowed fraction decreases with increasing the width of the mass function. We have identified three mass windows where an appreciable  fraction of DM can still consist of PBHs: $4\times10^{-17} \Msun$,  $2\times10^{-14} \Msun$ and $25 - 100 \Msun$. If all the constraints discussed in the literature are taken at face value and treated on an equal footing, then at most ${\cal O}(10\%)$ of DM can be in PBHs. However, if some of the dynamical constraints can be circumvented, then 100\% PBH DM might be allowed in these windows. Even ${\cal O}(10\%)$ DM in the ${\cal O}(10) \Msun$ window might suffice to explain the LIGO events.

%-------------------------------------------------------------------------------
\acknowledgments
We thank Juan Garcia-Bellido, Sebastien Clesse, Anne Green, Gert H\"utsi, Chloe James-Turner, Luca Marzola, Paul Schechter and Danton Weil for useful discussions. We also thank an anonymous referee for important input. This work was supported by the Estonian Research Council grants IUT23-6 and EU through the ERDF CoE program grant TK133. T.T. is supported by the U.K. Science and Technology Facilities Council grant ST/J001546/1.

%-------------------------------------------------------------------------------
\appendix*

\hypertarget{sec:combining}{}
\section{Combined constraints}
%-------------------------------------------------------------------------------

In this appendix we describe how different constraints can be combined. Consider $N$ independent observables $A_j(f_{\rm PBH})$ with observed expectation values $\mu_j$ and variances $\sigma_j^2$. The $\chi^2$ for these observables is
\be
\chi^2(f_{\rm PBH}) = \sum_{j=1}^N \frac{(A_j(f_{\rm PBH})-\mu_j)^2}{\sigma_j^2} \,.
\ee
We assume that the observables $A_j(f_{\rm PBH})$ are linear in the mass function, so that only the first two terms in~\eqref{eq:A_observable} are relevant, and that the mean values coincide with $f_{\rm PBH}=0$, which implies $\mu_j = A_{0,j}$. Since $\psi\propto f_{\rm PBH}$, the $n\sigma$ constraint on $f_{\rm PBH}$ is then
\be \label{eq:nsigmaconstraint}
n^2 \geq \chi^2-\chi_{\rm min}^2 = \sum_{j=1}^N \left(\int \td M \psi(M) \frac{K_{1,j}(M)}{\sigma_j} \right)^2 \,,
\ee
where $\chi_{\rm min}$ is the minimum of $\chi$. As in Eq.~\eqref{eq:f_max}, the kernel $K_{1,j}$ can be extracted from the constraint for a monochromatic mass function if $N=1$. This corresponds to
\be
f_{\rm PBH}(M) \leq \frac{n\sigma_j}{K_{1,j}(M)} \equiv f_{{\rm max},j}(M) \,.
\ee
It follows that Eq.~\eqref{eq:nsigmaconstraint} can be recast as
\be \label{A4eq}
\sum_{j=1}^N \left(\int \td M \frac{\psi(M)}{f_{{\rm max},j}(M)} \right)^2 \leq 1 \,.
\ee
Note that the upper bound in Eq.~\eqref{eq:Aexp} is $A_{{\rm exp},j} = A_{0,j} + n\sigma_j$, where $n\sigma_j$ is the confidence level of the bound. 

Consider then the combined constraint for a monochromatic mass function, which from Eq.~\eqref{A4eq} can be expressed as
\be\label{eq:fmax_combined}
	f_{\rm max}(M) = \left(\sum_{j=1}^N f_{{\rm max},j}(M)^{-2}\right)^{-1/2}\,.
\ee
Using this in Eq.~\eqref{eq:general_constraint} will always lead to an overestimation of the actual constraint given by \eqref{A4eq} because the triangle inequality implies 
\be
	\sum_{j=1}^N \left(\int \td M \frac{\psi(M)}{f_{{\rm max},j}(M)} \right)^2 \leq \left( \int \td M \frac{\psi(M)}{f_{\rm max}(M)} \right)^2 \,.
\ee 
For a more precise estimate, the contribution of each observable has to be included separately for any constraint derived from multiple observables. However, if the constraints $f_{{\rm max},j}(M)$ are proportional to each other, the expressions \eqref{A4eq} and \eqref{eq:general_constraint} are equivalent, so no error is made.

%-------------------------------------------------------------------------------
\bibliography{citations}

\end{document}